# Large tuning of the optical properties of nanoscale NdNiO$_3$ via electron doping

Yeonghoon Jin[1,#], Teng Qu[2,#], Siddharth Kumar[3,#], Nicola Kubzdela[2], Cheng-Chia Tsai[2], Tai De Li[4], Shriram Ramanathan[3,*], Nanfang Yu[2,*], Mikhail A. Kats[1,5,*]

[1]Department of Electrical and Computer Engineering, University of Wisconsin-Madison, Wisconsin 53706, USA

[2]Department of Applied Physics and Applied Mathematics, Columbia University, New York 10027, USA

[3]Department of Electrical and Computer Engineering, Rutgers, The State University of New Jersey, New Jersey 08854, USA

[4]CUNY Graduate Center Advanced Science Research Center, New York 11201, USA

[5]Department of Material Science and Engineering, University of Wisconsin-Madison, Wisconsin 53706, USA

[#]These authors contributed equally.

*Corresponding authors: Mikhail A. Kats (mkats@wisc.edu), Shriram Ramanathan (shriram.ramanathan@rutgers.edu), Nanfang Yu (ny2214@columbia.edu)

**Abstract**

We synthesized crystalline films of neodymium nickel oxide (NdNiO$_3$), a perovskite quantum material, switched the films from a metal phase (intrinsic) into an insulator phase (electron-doped) by field-driven lithium-ion intercalation, and characterized their structural and optical properties. Time-of-flight secondary-ion mass spectrometry (ToF-SIMS) showed that the intercalation process resulted in a gradient of the dopant concentration along the thickness direction of the films, turning the films into insulator–metal bilayers. We used variable-angle spectroscopic ellipsometry to measure the complex refractive indices of the metallic and insulating phases of NdNiO$_3$. The insulator phase has a refractive index of $n \sim 2$ and low absorption in the visible and near infrared, and analysis of the complex refractive indices indicated that the band gap of the insulating phase is roughly 3-4 eV. Electrical control of the optical band gap, with corresponding large changes to the optical refractive indices, creates new opportunities for tunable optics.

Phase-change materials (PCMs) with a reversible modulation of the complex refractive index—$n$ (real part) and/or $\kappa$ (imaginary part)—can enable tunable photonics applications from the visible to the terahertz.[1] Specific requirements for the optical properties of PCMs depend on the target applications, but an ideal PCM should offer large modulation of $n$, low loss (low $\kappa$), and fast switching speed over a wavelength range covering at least a portion of the visible or infrared (IR).[2] The realization of such tunable optical materials would benefit many domains of optics, from spatial light modulators[3] and displays[4], to lenses with tunable focal length[5] and other tunable free-space optics.[6] While many PCMs are being actively explored, including those based on crystalline-to-amorphous transitions such as Ge-Sb-Te (GST)[7] and Ge-Sb-Se-Te (GSST)[8], and insulator-to-metal transitions such as those in $VO_2$ and $V_2O_5$,[9] none have the combination of large tunability of $n$ and low $\kappa$ across a broad wavelength range in the visible and/or the near-infrared (near-IR). Therefore, there exists a need for new classes of tunable optical materials.

Perovskite rare-earth nickelates ($R$NiO$_3$, where $R$ is a rare-earth element such as Nd and Sm) have recently emerged as a class of PCMs that feature both temperature-driven and electron-doping-driven phase transitions. The thermal transition temperature, at which the materials change from a low-temperature insulating state to a high-temperature metallic state, depends on $R$, for example ~200 K for NdNiO$_3$ and ~400 K for SmNiO$_3$[10, 11]. However, these materials have substantial optical losses in both metallic (high-temperature) and semiconducting (low-temperature, with a small gap of the order of 0.1 eV) phases[12], and therefore applications of the thermally driven transition are limited; for example, SmNiO$_3$ has been used for dynamic tuning of thermal radiation[13], but is less suited for phase modulation in a waveguide or a resonant structure.

New opportunities for $R$NiO$_3$ have opened up with experiments demonstrating large and reversible tuning of the band gap via electric-field-driven electron doping at room temperature, achieved through the intercalation of H$^+$ or Li$^+$ ions[14]. First-principles theoretical calculations suggest that band gaps of ~3 eV can be opened via electron doping regardless of the ionic species used for intercalation[15, 16]. Experiments have demonstrated a change in electrical resistivity of 7–8 orders of magnitude[14, 17] and a transition from an optically lossy state to a low-loss state over a broad wavelength range from the visible to the IR[18]. Note that the doping-driven phase transition gives rise to a much larger change of material properties compared to those achievable using the temperature-driven phase transition,[11, 14] and that the electron-doping tuning is non-thermal, non-volatile, and controllable via electric fields.[18-20]

While several studies exist on the electrical resistivity changes due to electron doping, there have been very few studies on the optical properties of the intrinsic and electron-doped states of $R$NiO$_3$. A few experimental studies measured the complex refractive indices of intrinsic and electron-doped SmNiO$_3$,[14, 18] with the assumption that electron-doped SmNiO$_3$ films were uniformly doped. Several *ab-initio* studies have predicted the band gap of H$^+$- or Li$^+$-doped $R$NiO$_3$,[15, 16] but the results range from 0.5 to 5 eV, depending on the calculation method. As a result, there is a need for accurate measurement of the doping profile and the optical properties of electron-doped $R$NiO$_3$, and in particular, compounds like NdNiO$_3$, which are thermodynamically stable at atmospheric conditions and can be grown on many different substrates[20, 21].

Here, we measured the complex refractive indices of both undoped (metallic) and Li$^+$-doped (insulating) NdNiO$_3$ films using variable-angle spectroscopic ellipsometry (VASE) over a broad wavelength range from the ultraviolet to the near-IR ($\lambda$ = 0.3–2.5 μm). The films were electron-doped via intercalation of lithium (Li) ions using a liquid electrolyte from their top surfaces, and the doping process resulted in a non-uniform doping profile in the depth direction of the films, as characterized by time-of-flight secondary-ion mass spectrometry (ToF-SIMS). The films that are composed of a doped top layer and an under-doped bottom layer were characterized by VASE using a two-layer model. The results showed dramatically different complex refractive indices between the two phases, with the extinction coefficient (*i.e.*, imaginary part of the complex refractive index) $\kappa$ of Li$^+$-doped NdNiO$_3$ below 0.1 in the visible ($\lambda$ = 0.4–0.7 μm) and below 0.02 in the near-IR ($\lambda$ = 0.7–2.5 μm). Using these optical data, we estimated the band gap of the Li$^+$-doped NdNiO$_3$ to be roughly 3-4 eV.

NdNiO$_3$ crystalline thin films 50 nm in thickness were grown on 0.5% Nb-doped SrTiO$_3$ (Nb:STO) substrates via high-vacuum reactive magnetron sputtering at room temperature, as shown in **Figure 1a**. This substrate was chosen because of the small lattice mismatch with NdNiO$_3$. Two separate targets for Nd (125 W RF power) and Ni (75 W DC power) were co-sputtered in a mixed argon-oxygen (40:10 sccm) atmosphere maintained at 5 mTorr. During growth, the substrate holder was rotated at 25 rpm to ensure uniform growth. After deposition, the films were annealed at 500 ºC in ambient atmosphere for 24 h to achieve full oxidation of the films to the desired phase. The X-ray diffraction (XRD) measurement of the undoped NdNiO$_3$ on Nb:STO, shown in **Figure 1b**, confirms the small lattice mismatch and the epitaxial growth of NdNiO$_3$: the two adjacent peaks around 46.6° are from the crystalline Nb:STO substrate, and the single peak at 47.7° is from the undoped crystalline NdNiO$_3$ thin film; these peaks are very close to each other. Note that the Cu source used for the XRD measurement contains Cu Kα1, Kα2, and Kβ wavelengths, showing three peaks for Nb:STO.

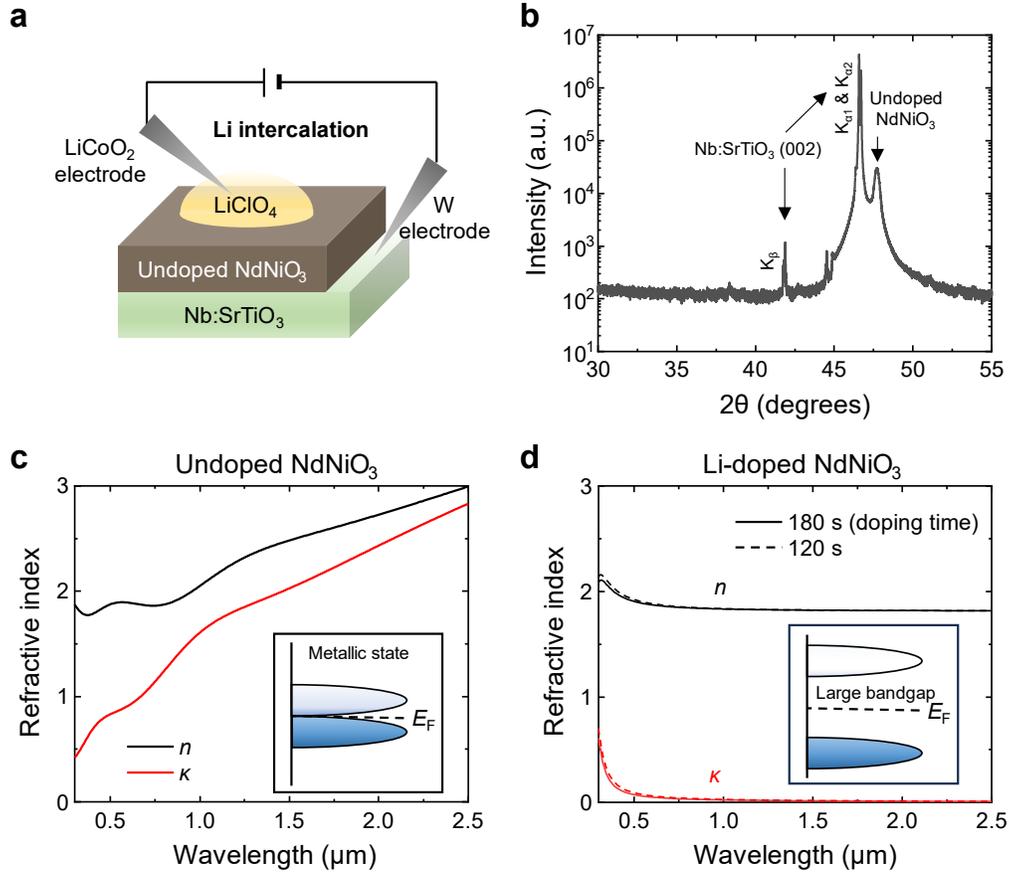

**Figure 1.** Structural and optical properties of the undoped and Li$^+$-doped NdNiO$_3$ thin films grown on Nb:STO substrates. **(a)** Depiction of the Li intercalation experiment. **(b)** X-ray diffraction (XRD) measurement of an undoped NdNiO$_3$/Nb:STO sample. **(c,d)** Measured complex refractive indices ($n$ and $\kappa$) of **(c)** the undoped NdNiO$_3$, which has no band gap at room temperature, and **(d)** the Li$^+$-doped NdNiO$_3$ films. The curves in (d) represent the complex refractive indices of the Li$^+$-doped films with different doping times of 120 and 180 s (refer to **Figure S1** for sample description).

To trigger the metal-to-insulator transition, a droplet of an electrolyte containing Li ions (0.1 mol saturated solution of LiClO$_4$ dissolved in propylene carbonate) was placed on the top of the film. Then, a top electrode (a LiCoO$_2$-coated aluminum foil) was dipped into the electrolyte, the other tungsten electrode was connected to the exposed bottom Nb:STO substrate (**Figure 1a**), and a positive voltage of 5.5 V was applied across the electrodes to drive Li ions into the NdNiO$_3$ thin film from the top electrode, while electrons were simultaneously driven into the thin film via the bottom electrode. The complex refractive index ($n$ and $\kappa$) of an undoped NdNiO$_3$ film is shown in **Figure 1c**, and that of Li$^+$-doped NdNiO$_3$ films with different doping times (120, and 180 s) is shown in **Figure 1d** (detailed explanation on how these results were obtained will be described in the next section). The complex refractive index of the undoped film is indicative of a metal, with $\kappa$ increasing as a function of wavelength, while that of the Li$^+$-doped films is indicative of an insulator possessing a large band gap, with small $\kappa$ decreasing with increasing wavelength. The average $\kappa$ in the visible ($\lambda$=0.4–0.7 µm) for the Li$^+$-doped state is lower than 0.1, and it decreases further in the near-IR.

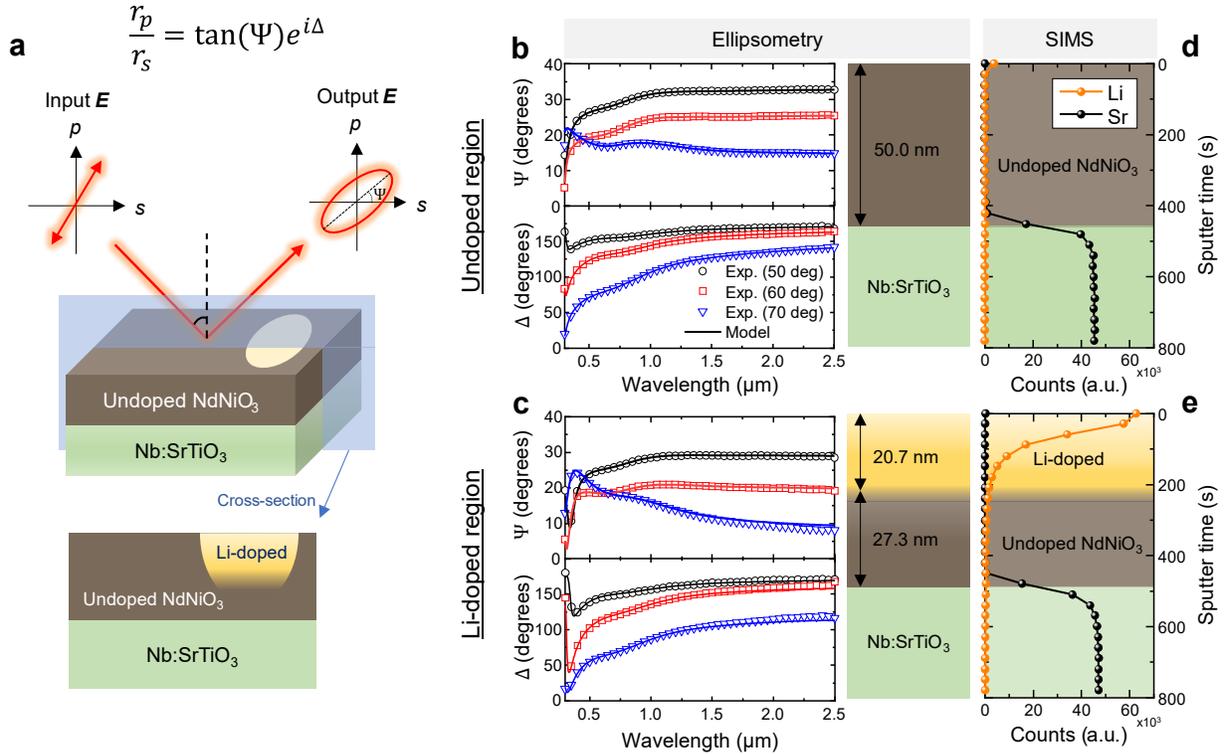

**Figure 2.** Spectroscopic ellipsometry and ToF-SIMS analysis of an NdNiO$_3$/Nb:STO sample. **(a)** Schematic of the ellipsometric measurement. There is a Li$^+$-doped region with a gradient Li ion concentration profile in the depth direction. **(b,c)** Ellipsometric parameters Ψ and Δ measured on **(b)** the undoped NdNiO$_3$ region and **(c)** the Li$^+$-doped NdNiO$_3$ region, at three angles of incidence of 50°, 60° and 70° (symbols). The solid curves represent a model fit, assuming one layer in the case of the undoped region and two layers for the doped region. The right panel depicts the estimated thickness of each layer based on ellipsometry. **(d,e)** Depth-dependent Li and Sr counts of **(d)** the undoped region and **(e)** the Li$^+$-doped region, measured by ToF-SIMS.

**Figure 2a** shows the process by which we extracted the complex refractive indices (shown in **Figures 1c** and **1d**) for both the doped and undoped NdNiO$_3$ regions. We used a VASE ellipsometer (J.A. Woollam Co.) in the wavelength range from 0.3 to 2.5 µm to measure the depolarization of light when it reflects from a NdNiO$_3$ film on an Nb:STO substrate. The output of a VASE measurement for a given angle of incidence are spectra of Ψ and Δ, defined as $\tan(\Psi) e^{i\Delta} = r_p/r_s$, where $r_p$ and $r_s$ are the complex-valued reflection coefficients for *p*- and *s*-polarized light, respectively.[22] By fitting the experimentally obtained wavelength-dependent Ψ and Δ with a model, we can determine the complex refractive index and thickness of the NdNiO$_3$ film.

We first measured the complex refractive index of the substrate (Nb:STO); see **Figure S2** for the ellipsometry measurements and resulting optical properties of Nb:STO. **Figure 2b** shows the parameters Ψ and Δ of the undoped NdNiO$_3$ region from λ = 0.3 to 2.5 µm, at three angles of incidence (50, 60, and 70 degrees). Note that a focusing probe was used to ensure that the beam spot was much smaller than the measured undoped NdNiO$_3$ region. The solid curves represent fitting of the experimental results using a single-layer model, showing good agreement between the model and the experiment. The NdNiO$_3$ complex refractive index is modeled using one Drude and three Lorentz oscillators (see **Note S1** in the Supporting Information for detailed description of the model). The fitted complex refractive index of undoped NdNiO$_3$ is shown in **Figure 1c** and the fitted thickness is 50 nm.

**Figure 2c** shows the parameters Ψ and Δ of the Li$^+$-doped region (doping time of 180 s) from λ = 0.3 to 2.5 µm, at three angles of incidence (50, 60, and 70 degrees). We were not able to fit the experimental Ψ and Δ spectra with a single-layer model, suggesting that the doped NdNiO$_3$ region is not uniform through its thickness. Because the source of Li ions was on the top surface of the film, it is reasonable to assume a gradient of the Li ion concentration in the depth direction. Thus, we used a two-layer model to fit the ellipsometry results, assuming that there is a fully doped layer on the top and an undoped layer on the bottom (lower panel in **Figure 2a**). Our two-layer model shows good agreement with the experimental Ψ and Δ (**Figure 2c**). The estimated thickness of the Li$^+$-doped NdNiO$_3$ layer is 20.7 nm and that of the undoped layer is 27.3 nm; the total thickness of 20.7+27.3 = 48 nm is close to the thickness measured in the undoped region of 50 nm. See **Note S1** for a detailed description of the two-layer model, the fitting process, and the oscillator parameters.

To validate our ellipsometry fitting model, we conducted ToF-SIMS measurements to characterize the vertical doping profile of Li ions (see **Note S2** for a description for the measurement conditions). **Figure 2e** shows the ToF-SIMS profile of Li and Sr, measured on the Li$^+$-doped region, and the same measurement was conducted on the undoped region for comparison (**Figure 2d**). The results show that the Li ion concentration is negligible in the undoped region, whereas in the Li$^+$-doped region the Li ion concentration exponentially decays from the top surface. Recent theoretical studies[23] have shown that the band gap rapidly opens up when the full electron doping level of one ion and one electron per Ni is approached (*e.g.*, in H$^+$-doped SmNiO$_3$, the band gap is only 1.4 eV at the half doping level of H$^+$:Ni=0.5:1, but abruptly increases to 5.1 eV at the full doping level of H$^+$:Ni=1:1). This justifies the two-layer model as an approximation of a partially doped thin film, where we assume that the fully doped and over-doped top layer has a fully opened band gap represented by *n* and *κ* characteristic of an insulator, and the undoped and insufficiently doped bottom layer has minimal band gap opening represented by *n* and *κ* that can be approximated by the complex refractive index of the undoped NdNiO$_3$.

We have also grown NdNiO$_3$ thin films on crystalline LaAlO$_3$ (LAO) substrates. This substrate was chosen because it provides a nearly lattice matched platform for epitaxial growth of crystalline NdNiO$_3$ and allows the change in transparency of Li$^+$-doped NdNiO$_3$ regions to be readily observed because LAO is transparent (unlike Nb:STO, which is opaque in the visible spectrum). Refer to **Figure S3** for X-ray diffraction (XRD) and Rutherford backscattering (RBS) characterization of the NdNiO$_3$/LAO sample. Since the LAO substrate is not electrically conductive, the voltage for Li intercalation of NdNiO$_3$ was applied between the top electrode in the electrolyte (as shown in **Figure 1a**) and a bottom electrode placed on undoped NdNiO$_3$, with 5.5 V applied across the electrodes for 10 minutes. **Figure 3a** shows a photo of the sample after the Li intercalation experiment: the large transparent region is where NdNiO$_3$ is doped with Li ions, the dark region represents undoped NdNiO$_3$, and the smallest region is the exposed LAO substrate, from where the sample was mounted using a clip in the sputtering chamber. The doped region is nearly as transparent as the exposed substrate, suggesting low *κ* of the doped NdNiO$_3$ in the visible.

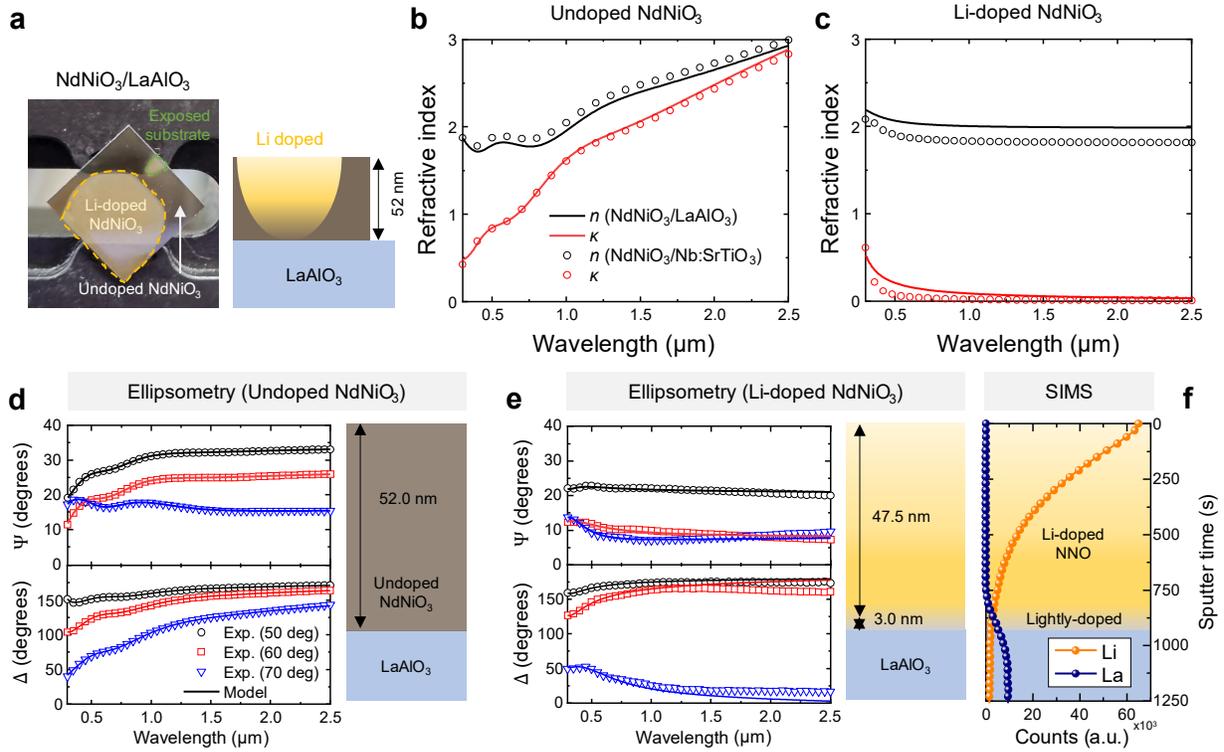

**Figure 3.** Measurement of undoped and Li$^+$-doped NdNiO$_3$ thin films grown on LaAlO$_3$ (LAO) substrates. **(a)** A photo of our NdNiO$_3$/LAO sample, including the Li$^+$-doped region, the undoped region, and the exposed substrate region. **(b,c)** Extracted complex refractive indices of **(b)** the undoped NdNiO$_3$ and **(c)** the Li$^+$-doped NdNiO$_3$ (curves). The complex refractive indices of NdNiO$_3$ on Nb:STO (extracted from Figures 2b and 2c) are also shown for comparison (symbols). **(d,e)** The Ψ and Δ data of **(d)** the undoped NdNiO$_3$ and **(e)** the Li$^+$-doped NdNiO$_3$, at three angles of incidence (50°, 60°, and 70°). The solid curves represent a model fit. The right panel shows the thickness estimated by ellipsometry. **(f)** ToF-SIMS results of the doped NdNiO$_3$ region, showing the Li and La concentration depth profiles.

VASE and ToF-SIMS were used to characterize the NdNiO$_3$/LAO samples as well as the LAO substrate. The ellipsometry data and the refractive index of the LAO substrate are shown in **Figure S4**. The refractive index of the undoped NdNiO$_3$ region on LAO are shown in **Figure 3b** and that of the Li$^+$-doped NdNiO$_3$ region are shown in **Figure 3c**, both very similar to the measurements of the NdNiO$_3$ films on Nb:STO. The thickness of the undoped NdNiO$_3$ film on LAO extracted from VASE is 52 nm (**Figure 3d**). For the doped region, the two-layer model shows good fitting with the experimental data (**Figure 3e**), and the estimated thickness of the top layer is 47.5 nm and that of the bottom layer is only 3 nm (refer to **Note S1** for a detailed description of the fitting process). ToF-SIMS independently confirmed that Li ions are present almost over the entire thickness of the NdNiO$_3$ film with a gradient concentration profile (**Figure 3f**). The deeper Li ion penetration depth for the NdNiO$_3$/LAO sample results from a longer Li intercalation time (600 s) compared to that of the NdNiO$_3$/Nb:STO sample in **Figure 2e** (60–180 s).

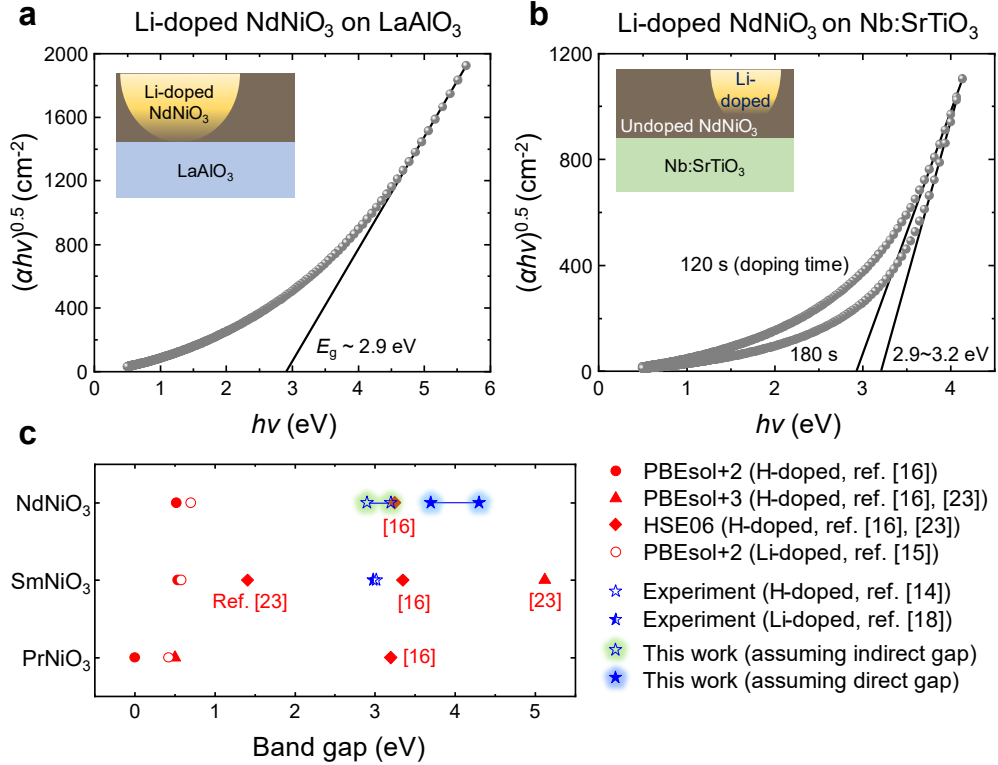

**Figure 4.** Tauc plots to estimate the optical band gap of doped NdNiO$_3$, assuming that the gap is indirect. **(a)** Tauc plot for the Li$^+$-doped NdNiO$_3$ film on LAO, calculated from the data in Figure 3c. **(b)** The same plot for the Li$^+$-doped NdNiO$_3$ film on Nb:STO, calculated from the data in Figure 1d. Based on these plots, the estimated band gap is roughly 3-4 eV. **(c)** Theoretically and experimentally estimated band gaps of H$^+$- or Li$^+$-doped PrNiO$_3$, SmNiO$_3$, and NdNiO$_3$ from the literature[14-16, 18, 23] and from our experiments, with two assumptions: (i) assuming the gap is indirect, as in (a, b), and (ii) assuming that the gap is direct, as shown in Figure S6.

A few theoretical papers have made first-principles predictions of the band gap of H$^+$- or Li$^+$-doped $R$NiO$_3$[15, 16], but the predicted band gaps vary over a wide range due to the assumptions used in the calculations. Here, we estimated the band gap of Li$^+$-doped NdNiO$_3$ using the Tauc plot method. The Tauc plot equation is given by: $(\alpha h\nu)^m = A(h\nu - E_g)$, where $\alpha$ is the absorption coefficient, $h$ is the Planck constant, $\nu$ is the frequency of incident light, $A$ is a constant, and $E_g$ is the optical band gap. Extrapolating the linear region of the Tauc plot to the horizontal axis yields the band gap.[24] The power, $m$, is 0.5 for indirect transitions and 2 for direct transitions[24].

**Figures 4a,b** show the Tauc plots of the Li$^+$-doped NdNiO$_3$ films on LAO and Nb:STO, respectively, assuming an indirect gap. The estimated band gap is 2.9–3.2 eV, matching with some of the theoretical predictions[16] and experimental reports on H$^+$- or Li$^+$-doped SmNiO$_3$[14, 18]. Assuming that the gap is direct leads to an estimated gap of 3.7–4.3 eV (**Figure S6**). **Figure 4c** shows the theoretically[15, 16, 23] and experimentally[14, 18] estimated band gaps of H$^+$- or Li$^+$-doped PrNiO$_3$, SmNiO$_3$, and NdNiO$_3$, including our experiments. The band gap of H$^+$- or Li$^+$-doped $R$NiO$_3$ estimated by first-principles calculations varies from ~0 to ~5.1 eV depending on the underlying assumptions. A number of these theoretical results are aligned with our experimental estimates, which indicates the band gap of roughly 3-4 eV.

We note that a recent paper reported that the Tauc plot method may not be applicable to Mott-transition materials such as Co$_3$O$_4$ and Co$_x$Fe$_{3-x}$O$_4$ because the complex bands permit many optical transitions so it is not possible to determine which absorption is originated from the valence band maximum to the conduction band minimum.[24] However, unlike Co$_3$O$_4$ and Co$_x$Fe$_{3-x}$O$_4$, the optical absorption of our Li$^+$-doped NdNiO$_3$ is substantial only at wavelengths shorter than 400 nm (**Figure 1d**), which suggests that the absorption is caused by the band edge-to-edge transition. Thus, we expect that the Tauc plot can still be used to estimate the band gap of H$^+$- or Li$^+$-doped $R$NiO$_3$.

In summary, we characterized, for the first time to our knowledge, the complex refractive indices of both undoped and Li$^+$-doped NdNiO$_3$ films at wavelengths from 0.3 to 2.5 μm. Upon electron doping via Li intercalation, NdNiO$_3$ undergoes a phase transition, where the band gap changes from near-zero (metallic state) to 3-4 eV (insulating state). In our samples, the Li ion concentration profile in the NdNiO$_3$ films has a gradient in the depth direction, as evidenced by ToF-SIMS; thus, we introduced a two-layer model to fit data from VASE. The results show that the extinction coefficient ($\kappa$) of the insulating state of NdNiO$_3$ is lower than 0.1 in the visible and 0.02 in the near-IR. Future efforts to improve the performance of the rare-earth perovskite nickelates will focus on precisely controlling the electron doping levels via Li or proton intercalation so that we can leverage the large optical refractive index dispersion in proximity to the band edge to realize large and continuous refractive index tuning and low optical losses across the visible and near-IR ranges. Given the broad interest of tuning physical properties of quantum materials via electric-field gating, the method presented here to analyze nanoscale matter with inhomogeneous electron concentrations and the large tuning of optical properties should open new directions to design electrically tunable solids.

## Acknowledgements

The authors acknowledge helpful discussions with Rohith Chandrasekar, Trish Veeder, Arash Dehzangi, Jonathan Slagle, Michael Carter, and Yuan Yang. The work was supported by the DARPA ATOM Program (HR00112390123)

## Conflict of Interest

The authors declare no conflict of interests.

## References

(1) Ko, J. H.; Yoo, Y. J.; Lee, Y.; Jeong, H.-H.; Song, Y. M. A review of tunable photonics: Optically active materials and applications from visible to terahertz. *IScience* **2022**, *25* (8).

(2) Veeder, T.; Dehzangi, A.; Ramanathan, S.; Kats, M.; Yu, N.; Hu, J.; Roberts, C.; Polking, M.; Tibbetts, K.; Majumdar, A.; et al. Accelerating discovery of tunable optical materials (ATOM). In *Image Sensing Technologies: Materials, Devices, Systems, and Applications XI*, 2024; SPIE: Vol. 13030, p 1303002.

(3) Li, S.-Q.; Xu, X.; Maruthiyodan Veetil, R.; Valuckas, V.; Paniagua-Domínguez, R.; Kuznetsov, A. I. Phase-only transmissive spatial light modulator based on tunable dielectric metasurface. *Science* **2019**, *364* (6445), 1087-1090.

(4) Hosseini, P.; Wright, C. D.; Bhaskaran, H. An optoelectronic framework enabled by low-dimensional phase-change films. *Nature* **2014**, *511* (7508), 206-211.

(5) Wang, Q.; Rogers, E. T.; Gholipour, B.; Wang, C.-M.; Yuan, G.; Teng, J.; Zheludev, N. I. Optically reconfigurable metasurfaces and photonic devices based on phase change materials. *Nature Photonics* **2016**, *10* (1), 60-65.

(6) Julian, M. N.; Williams, C.; Borg, S.; Bartram, S.; Kim, H. J. Reversible optical tuning of GeSbTe phase-change metasurface spectral filters for mid-wave infrared imaging. *Optica* **2020**, *7* (7), 746-754.

(7) Wuttig, M.; Bhaskaran, H.; Taubner, T. Phase-change materials for non-volatile photonic applications. *Nature Photonics* **2017**, *11* (8), 465-476.

(8) Zhang, Y.; Chou, J. B.; Li, J.; Li, H.; Du, Q.; Yadav, A.; Zhou, S.; Shalaginov, M. Y.; Fang, Z.; Zhong, H.; et al. Broadband transparent optical phase change materials for high-performance nonvolatile photonics. *Nature Communications* **2019**, *10* (1), 4279.

(9) Ko, B.; Badloe, T.; Rho, J. Vanadium dioxide for dynamically tunable photonics. *ChemNanoMat* **2021**, *7* (7), 713-727.

(10) Lacorre, P.; Torrance, J.; Pannetier, J.; Nazzal, A.; Wang, P.; Huang, T. Synthesis, crystal structure, and properties of metallic PrNiO3: Comparison with metallic NdNiO3 and semiconducting SmNiO3. *Journal of Solid State Chemistry* **1991**, *91* (2), 225-237.

(11) Torrance, J.; Lacorre, P.; Nazzal, A.; Ansaldo, E.; Niedermayer, C. Systematic study of insulator-metal transitions in perovskites R NiO 3 (R= Pr, Nd, Sm, Eu) due to closing of charge-transfer gap. *Physical Review B* **1992**, *45* (14), 8209.

(12) Stewart, M.; Liu, J.; Kareev, M.; Chakhalian, J.; Basov, D. Mott physics near the insulator-to-metal transition in NdNiO 3. *Physical Review Letters* **2011**, *107* (17), 176401.

(13) Shahsafi, A.; Roney, P.; Zhou, Y.; Zhang, Z.; Xiao, Y.; Wan, C.; Wambold, R.; Salman, J.; Yu, Z.; Li, J.; et al. Temperature-independent thermal radiation. *Proceedings of the National Academy of Sciences* **2019**, *116* (52), 26402-26406.

(14) Shi, J.; Zhou, Y.; Ramanathan, S. Colossal resistance switching and band gap modulation in a perovskite nickelate by electron doping. *Nature Communications* **2014**, *5* (1), 4860.

(15) Cui, Y.; Liu, X.; Fan, W.; Ren, J.; Gao, Y. Metal–insulator transition in RNiO3 (R= Pr, Nd, Sm, Gd, Tb, Dy, Ho, Er) induced by Li doping: A first-principles study. *Journal of Applied Physics* **2021**, *129* (23).

(16) Yoo, P.; Liao, P. First principles study on hydrogen doping induced metal-to-insulator transition in rare earth nickelates RNiO3 (R= Pr, Nd, Sm, Eu, Gd, Tb, Dy, Yb). *Physical Chemistry Chemical Physics* **2020**, *22* (13), 6888-6895.

(17) Sun, Y.; Kotiuga, M.; Lim, D.; Narayanan, B.; Cherukara, M.; Zhang, Z.; Dong, Y.; Kou, R.; Sun, C.-J.; Lu, Q.; et al. Strongly correlated perovskite lithium ion shuttles. *Proceedings of the National Academy of Sciences* **2018**, *115* (39), 9672-9677.

(18) Li, Z.; Zhou, Y.; Qi, H.; Pan, Q.; Zhang, Z.; Shi, N. N.; Lu, M.; Stein, A.; Li, C. Y.; Ramanathan, S.; et al. Correlated Perovskites as a New Platform for Super-Broadband-Tunable Photonics. *Advanced Materials* **2016**, *28* (41), 9117-9125.

(19) Shi, J.; Ha, S. D.; Zhou, Y.; Schoofs, F.; Ramanathan, S. A correlated nickelate synaptic transistor. *Nature Communications* **2013**, *4* (1), 2676.

(20) Zhang, H.-T.; Park, T. J.; Islam, A. N.; Tran, D. S.; Manna, S.; Wang, Q.; Mondal, S.; Yu, H.; Banik, S.; Cheng, S.; et al. Reconfigurable perovskite nickelate electronics for artificial intelligence. *Science* **2022**, *375* (6580), 533-539.

(21) Hauser, A. J.; Mikheev, E.; Moreno, N. E.; Hwang, J.; Zhang, J. Y.; Stemmer, S. Correlation between stoichiometry, strain, and metal-insulator transitions of NdNiO3 films. *Applied Physics Letters* **2015**, *106* (9).

(22) Fujiwara, H.; Collins, R. W. *Spectroscopic ellipsometry for photovoltaics*; Springer, 2018.

(23) Yamauchi, K.; Hamada, I. Hydrogen-induced insulating state accompanied by interlayer charge ordering in SmNiO 3. *Physical Review B* **2023**, *108* (4), 045108.

(24) Klein, J.; Kampermann, L.; Mockenhaupt, B.; Behrens, M.; Strunk, J.; Bacher, G. Limitations of the Tauc plot method. *Advanced Functional Materials* **2023**, *33* (47), 2304523.

# Supporting information:

# Large tuning of the optical properties of nanoscale NdNiO$_3$ via electron doping

Yeonghoon Jin[1,#], Teng Qu[2,#], Siddharth Kumar[3,#], Nicola Kubzdela[2], Cheng-Chia Tsai[2], Tai De Li[4], Shriram Ramanathan[3,*], Nanfang Yu[2,*], Mikhail A. Kats[1,5,*]

[1]Department of Electrical and Computer Engineering, University of Wisconsin-Madison, Wisconsin 53706, USA

[2]Department of Applied Physics and Applied Mathematics, Columbia University, New York 10027, USA

[3]Department of Electrical and Computer Engineering, Rutgers, The State University of New Jersey, New Jersey 08854, USA

[4]CUNY Graduate Center Advanced Science Research Center, New York 11201, USA

[5]Department of Material Science and Engineering, University of Wisconsin-Madison, Wisconsin 53706, USA

[#]These authors contributed equally.

*Corresponding authors: Mikhail A. Kats (mkats@wisc.edu), Shriram Ramanathan (shriram.ramanathan@rutgers.edu), Nanfang Yu (ny2214@columbia.edu)

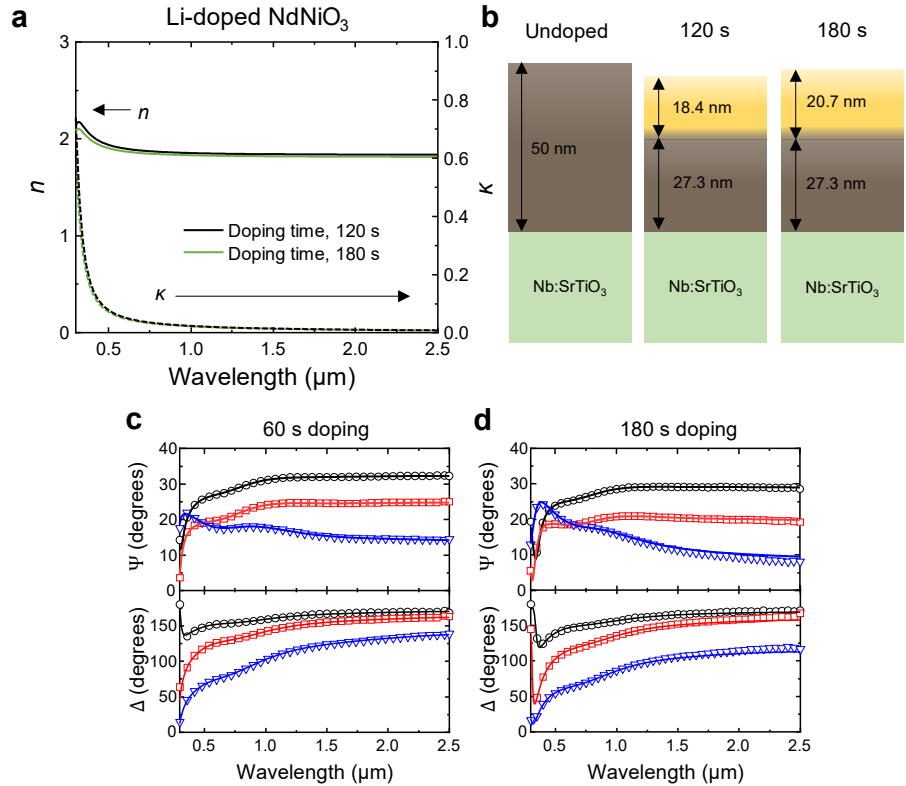

**Figure S1.** Optical properties of two Li-doped NdNiO$_3$ films on Nb:STO (0.5% Nb-doped SrTiO$_3$) with different doping times. **(a)** Complex refractive index of the Li-doped NdNiO$_3$ with doping times of 120 and 180 s. **(b)** Illustrations of the samples with the thickness estimated by ellipsometry. Doping was conducted on one chip but at different spots, and therefore the entire film thickness of each spot should be the same as in the undoped region (50 nm). However, the film thicknesses of the doped spots are slightly off from 51 nm, and this discrepancy might be because we used a two-layer model (the top doped and bottom undoped layers) without considering intermediate doping states between the top and bottom layers. Nevertheless, the independently obtained refractive indices of the Li-doped NdNiO$_3$ are consistent with each other, as shown in (a), and we believe the results are reliable. **(c-e)** Ellipsometry results (Ψ and Δ) of the two spots with different doping time. See **Note S1** for detailed description for the oscillators used to fit the experiments.

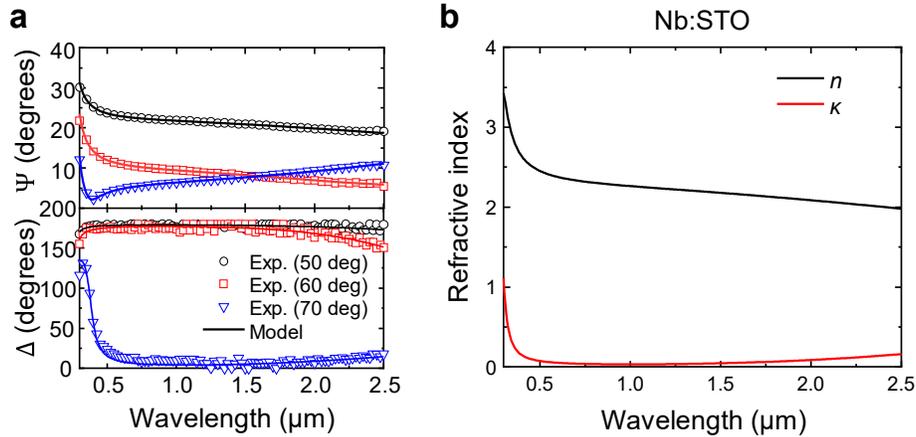

**Figure S2.** Optical properties of 0.5% Nb-doped $SrTiO_3$ (Nb:STO). **(a)** Spectroscopic ellipsometry results (Ψ and Δ) at three angles of incidence (50, 60, and 70 degrees) with the model fits. The oscillators used in the model are summarized in **Table S1**. **(b)** Complex refractive index of Nb:STO.

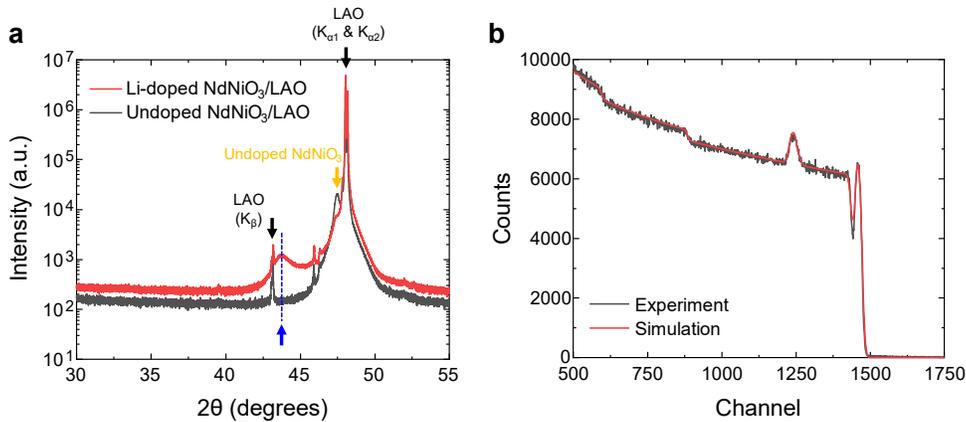

**Figure S3.** Material characterization for $NdNiO_3$ film on $LaAlO_3$ (LAO). **(a)** X-ray diffraction (XRD) data for both the undoped and Li-doped $NdNiO_3$ regions. The peak at 47.5° represents the undoped $NdNiO_3$ and the two adjacent peaks around 48.1° represents the LAO substrate; the $NdNiO_3$ and LAO peaks are very close each other, indicating the small lattice mismatch. After Li doping, a reduced intensity of the undoped $NdNiO_3$ peak (at 47.5°, indicated by the yellow arrow) is observed, along with the appearance of a feature at lower 2θ values (at 43.7°, indicated by the blue arrow). This is very similar to the case of H-doped $NdNiO_3$, where the absorbed H-ions get bonded to the O-ions in $NiO_6$ octahedra, increasing the lattice constant and opening a bandgap because of electron transfer.[1] Note that the Cu source used in the measurements was not monochromatic, and contained Cu-$K\alpha_1$, $K\alpha_2$, and $K\beta$ wavelengths. **(b)** Rutherford backscattering spectroscopy (RBS) data for the undoped $NdNiO_3$/LAO, obtained by bombarding a 2-mm-wide 2.3 MeV $He^{2+}$ ion beam and observing the backscattered particles. The x-axis (channels) represents the energy of the backscattered particles: the higher the channel number, the heavier the particle. The simulation (the red line) was conducted by assuming the $NdNiO_3$ film composition of Nd:Ni:O = 22:18:60 and the film thickness

of 38 nm, and it shows good agreement with the experiment. This confirms that the film is uniform throughout its thickness.

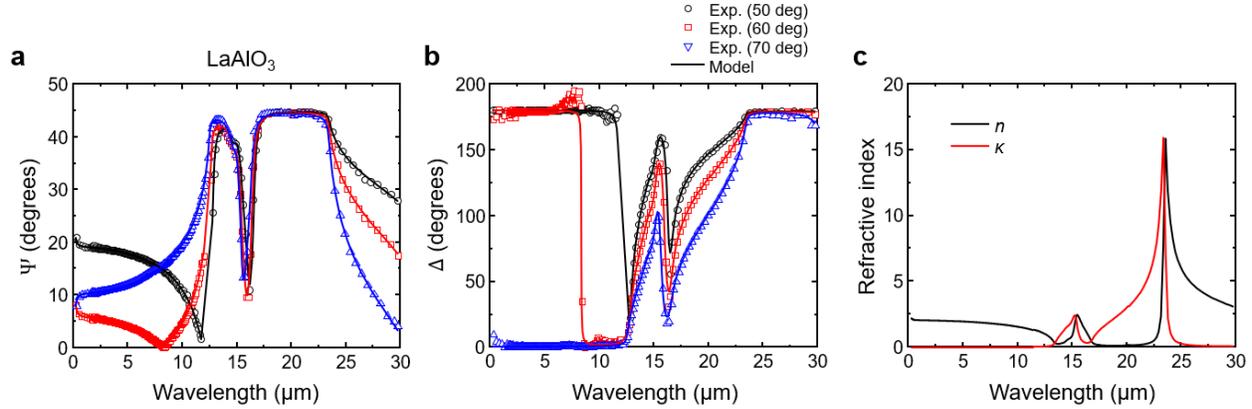

**Figure S4.** Optical properties of LaAlO$_3$ (LAO). **(a)** Spectroscopic ellipsometry results ($\Psi$ and $\Delta$) at three angles of incidence (50, 60, and 70 degrees) with the model fits. The oscillators used in the model are summarized in **Table S1**. **(b)** Complex refractive index of LAO.

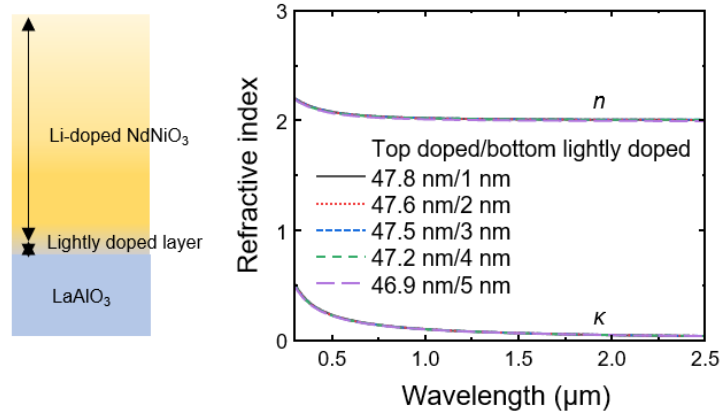

**Figure S5.** The refractive indices of the Li-doped NdNiO$_3$ on a LaAlO$_3$ substrate obtained from the ellipsometry results (**Figure 3e** in the main text), with different thickness combinations of the top doped and bottom lightly doped NdNiO$_3$. For the ellipsometry data analysis, we enforced the thickness of the bottom lightly doped layer from 1 to 5 nm, with intervals of 1 nm, and therefore, there were three unknown parameters: the refractive index of the top Li-doped and bottom lightly doped layers, and the thickness of the top layer. The refractive index of the Li-doped NdNiO$_3$ is rarely affected by the thickness of the lightly doped layer.

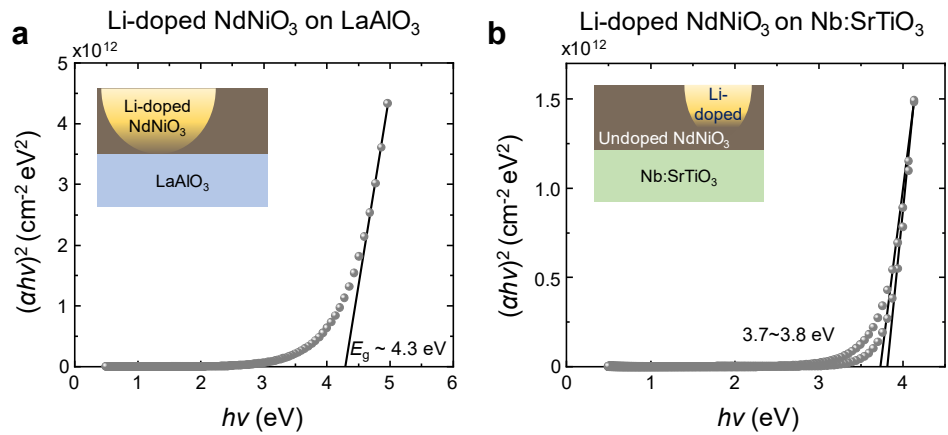

**Figure S6.** Tauc plots with an assumption that Li-doped NdNiO$_3$ is a direct-gap material. **(a,b)** Tauc plot of Li-doped NdNiO$_3$ on **(a)** LaAlO$_3$ and **(b)** Nb:SrTiO$_3$.

**Note S1.** Oscillators for the analysis of ellipsometry results

<u>Drude oscillator</u>

Drude oscillator can be given by:[2]

$$\varepsilon(E) = \varepsilon_1 - i\varepsilon_2 = \varepsilon_{1\infty} - \frac{A_n Br_n}{E^2 - iBr_n E}. \tag{S1}$$

$\varepsilon_1$ and $\varepsilon_2$ are the real and imaginary parts of complex permittivity, $\varepsilon_\infty$ is the high-frequency dielectric constant, $E$ is the photon energy in eV, $A_n$ is the oscillator strength of $n^{th}$ oscillator, and $Br_n$ is the broadening factor.

<u>Lorentz oscillator</u>

Lorentz oscillators can be given by:[2]

$$\varepsilon(E) = \varepsilon_1 - i\varepsilon_2 = \varepsilon_{1\infty} + \frac{A_n E_n}{E_n^2 - E^2 + iBr_n E}. \tag{S2}$$

$\varepsilon_1$ and $\varepsilon_2$ are the real and imaginary parts of complex permittivity, $\varepsilon_\infty$ is the high-frequency dielectric constant, $E_n$ is the center energy of $n^{th}$ oscillator, $E$ is the photon energy in eV, $A_n$ is the oscillator strength, and $Br_n$ is the broadening factor.

<u>Gaussian oscillator</u>

Gaussian oscillators can be given by:[2]

$$\varepsilon_2(E) = A_n/Br_n e^{-\left(\frac{E-E_n}{\sigma}\right)^2} - A_n/Br_n e^{-\left(\frac{E+E_n}{\sigma}\right)^2}, \tag{S3}$$

where $A_n$ is the oscillator strength of $n^{th}$ oscillator, $E$ is the energy in eV, $E_n$ is the center energy, and $\sigma = Br_n/2\sqrt{\ln(2)}$.[2] $\varepsilon_1$ can be obtained by Kramers-Kronig relations, which is given by:

$$\varepsilon_1(E) = 1 + \frac{2}{\pi} P \int_0^\infty \frac{E' \varepsilon_2(E')}{E'^2 - E^2} dE', \tag{S4}$$

where $P$ is the principal part of the integration.

<u>Cauchy</u>

Cauchy equation can be given by:[2]

$$n(\lambda) = A_n + \frac{B_n}{\lambda^2} + \frac{C_n}{\lambda^4}. \tag{S5}$$

$A_n$ represents the long-wavelength index, $B_n$ and $C_n$ are the dispersion terms. This model assumes that the material is lossless, and therefore, $\varepsilon(E) = \varepsilon_1(E) = n^2$.

## Two-layer model (NdNiO$_3$ on Nb:STO, Figure 2 in the main text)

The refractive index of the undoped NdNiO$_3$ (on Nb:STO) and the total film thickness (50 nm) were obtained by the ellipsometer measurement on the undoped NdNiO$_3$ region, as shown in **Figure 2b** in the main text. A two-layer model was used to fit the ellipsometer data for the Li-doped NdNiO$_3$ region: the top fully doped NdNiO$_3$ and the bottom undoped NdNiO$_3$ (**Figure 2c**). There are two remaining unknowns at this region: the thickness the Li-doped layer (the thickness of the undoped layer is known because the total thickness of 50 nm was obtained from the undoped region), and the refractive index of the Li-doped NdNiO$_3$. Detailed oscillators are summarized in **Table S1**.

## Two-layer model (NdNiO$_3$ on LaAlO$_3$, Figure 3 in the main text)

Based on the ToF-SIMS measurement (**Figure 3f** in the main text) and the degree of visible transparency we observe, we assumed that the thickness of the bottom lightly doped layer is most likely smaller than 5 nm. We then performed 5 separate fits, each assuming a different thickness of the bottom layer from 1 to 5 nm. Each of these fits had three fitting parameters: the refractive index of the top and bottom NdNiO$_3$ layers, and the thickness of the top Li-doped layer. We then found the best of the 5 fits, with the result shown in **Figure 3e**. The estimated thickness in this region (47.5 + 3 nm = 50.5 nm) almost matches the overall thickness of 52 nm. Refer to **Figure S5** for the refractive index values of the top doped NdNiO$_3$ layer obtained by the five separate fits.

**Table S1.** Summary of oscillator types and values used in our work.

| Samples | # | Oscillators | $A_n$ (eV) | $E_n$ (eV) | $Br_n$ (eV) |
|---|---|---|---|---|---|
| **NdNiO$_3$ on Nb:STO** | | | | | |
| **Substrate (Nb:STO)** | | | $\varepsilon_{1\infty}$ = 2.1281 | | |
| | 1 | Drude | 1.6942 | | 0.21693 |
| | 2 | Gaussian | 0.13904 | 2.2568 | 1.878 |
| | 3 | Gaussian | 2.6945 | 4.0872 | 0.52314 |
| | 4 | Gaussian | 5.1596 | 4.7317 | 1.1368 |
| | 5 | Gaussian | 4.5801 | 6.4149 | 0.21157 |
| Undoped NdNiO$_3$ | | | $\varepsilon_{1\infty}$ = 2.8872 | | |
| | 1 | Drude | 9.9825 | | 1.0518 |
| | 2 | Lorentz | 5.1795 | 5.2621 | 1.1724 |
| | 3 | Lorentz | 5.1123 | 2.7884 | 2.6075 |
| | 4 | Lorentz | 2.8448 | 1.1441 | 1.0536 |
| | 1 | Lorentz | 13.23 | 7.259 | 5.9142 |

| | | | $\varepsilon_{1\infty}$ = 2.3774 | | |
|---|---|---|---|---|---|
| Doped NdNiO$_3$ (120 s) | 1 | Lorentz | 4.3692 | 4.4246 | 1.1855 |
| Doped NdNiO$_3$ (180 s) | | | $\varepsilon_{1\infty}$ = 2.6308 | | |
| | 1 | Lorentz | 3.3265 | 4.3961 | 1.0661 |
| | | | NdNiO$_3$ on LAO | | |
| Substrate (LAO) | | | $\varepsilon_{1\infty}$ = 3.9574 | | |
| | 1 | Cauchy | $A_n$ = 0.39829, $B_n$ = 0.043615 (these values are dimensionless) | | |
| | 2 | Lorentz | 0.35047 | 0.02257 (182.0 cm$^{-1}$) | 0.33821 x 10$^{-3}$ |
| | 3 | Lorentz | 0.22239 | 0.05287 (426.4 cm$^{-1}$) | 0.55318 x 10$^{-3}$ |
| | 4 | Lorentz | 0.22878 x 10$^{-3}$ | 0.06148 (495.8 cm$^{-1}$) | 0.47724 x 10$^{-3}$ |
| | 5 | Lorentz | 0.022063 | 0.08073 (651.1 cm$^{-1}$) | 2.4718 x 10$^{-3}$ |
| | 6 | Lorentz | 4.0938 x 10$^{-3}$ | 0.08446 (681.1 cm$^{-1}$) | 4.4059 x 10$^{-3}$ |
| Undoped NdNiO$_3$ | | | $\varepsilon_{1\infty}$ = 1.8425 | | |
| | 1 | Drude | 10.092 | | 0.9752 |
| | 2 | Lorentz | 13.32 | 6.8453 | 2.6173 |
| | 3 | Lorentz | 3.7493 | 2.6041 | 2.1559 |
| | 4 | Lorentz | 2.8842 | 1.1241 | 1.0388 |
| Doped NdNiO$_3$ | | | $\varepsilon_{1\infty}$ = 1.0729 | | |
| | 1 | Lorentz | 4.8415 | 6.3222 | 3.7405 |
| Lightly doped layer | | | $\varepsilon_{1\infty}$ = 9.3364 | | |
| | 1 | Lorentz | 95.653 | 0.94989 | 10 |

**Note S2.** Time-of-flight (ToF) secondary ion mass spectrometry (SIMS)

An Ar$^+$ beam (20 kV, 500 nA) was used for milling of an area of 1 mm × 1 mm over a sample, and a Bi$^+$ beam (30 kV) was used for the primary analysis over a 100 μm × 100 μm zone located at the center of the milled area. During the measurements, a low-energy electron gun and a gas gun (Ar$^+$) were used to neutralize the surface charge.

**Supplementary References**

(1) Sidik, U.; Hattori, A. N.; Hattori, K.; Alaydrus, M.; Hamada, I.; Pamasi, L. N.; Tanaka, H. Tunable Proton Diffusion in NdNiO3 Thin Films under Regulated Lattice Strains. *ACS Applied Electronic Materials* **2022**, *4* (10), 4849-4856.
(2) Fujiwara, H.; Collins, R. W. *Spectroscopic ellipsometry for photovoltaics*; Springer, 2018.